\documentclass[11pt,a4paper]{article}
\usepackage[T1]{fontenc}
\usepackage[utf8]{inputenc}
\usepackage{lmodern}
\usepackage[margin=25mm]{geometry}
\usepackage{amsmath,amssymb,graphicx,booktabs,array}
\usepackage{authblk}
\usepackage[authoryear,round]{natbib}
\usepackage{xurl}
\usepackage[hidelinks]{hyperref}
\usepackage{microtype}
\providecommand{\doi}[1]{\href{https://doi.org/#1}{\nolinkurl{#1}}}
\title{Improved Myelin Water Fraction Estimation and Uncertainty Mapping in Multicomponent \texorpdfstring{$T_2$}{T2} Relaxometry via NNLS Bootstrap Aggregation}

\author[1,2,3]{Erick J Canales-Rodríguez}

\author[4]{Thomas Yu}
\author[1,2,3] {Juan Manuel Górriz}
\author[5,6,7]{Santiago Galdo-Álvarez}
\author[8]{Tobias Kober}
\author[4]{Tom Hilbert}
\author[4]{Gian Franco Piredda}
\author[9]{Miguel Ángel Rivas-Fernández} 

\affil[1]{{Department of Signal Theory, Networking and Communications, University of Granada},
            {Granada},
            {18071}, 
            {Spain}}

\affil[2]{{Research Centre for Information and Communication Technologies (CITIC), University of Granada},
            {Granada},
            {18071}, 
            {Spain}}

\affil[3]{{Andalusian Research Institute in Data Science and Computational Intelligence (DaSCI)},
            {Granada},
            {18016}, 
            {Spain}}

\affil[4]{{Swiss Innovation Hub, Siemens Healthineers International AG},
            {Lausanne},
            {CH-1015}, 
            {Switzerland}}

\affil[5]{{Department of Clinical Psychology and Psychobiology, University of Santiago de Compostela},
        {Santiago de Compostela},
            {15782}, 
            {Spain}}

\affil[6]{{Cognitive Neuroscience Research and Psychogerontology Group (NeuCogA-Aging), Institute of Psychology (IPsiUS), University of Santiago de Compostela},
            {Santiago de Compostela},
            {15782}, 
            {Spain}}
            
\affil[7]{{Health Research Institute of Santiago de Compostela},
            {Santiago de Compostela},
            {15706}, 
            {Spain}}

\affil[8]{{Research \& Clinical Translation, Magnetic Resonance, Siemens Healthineers AG},
            {Erlangen},
            {91052}, 
            {Germany}}
            
\affil[9]{{Department of Psychology, Sociology and Philosophy, University of León},
            {León},
            {24071}, 
            {Spain}}

\date{}
\begin{document}
\maketitle

\begin{abstract}
Estimating the myelin water fraction (MWF) from multicomponent \(T_2\) relaxometry requires solving an ill-conditioned inverse problem. As a result, MWF estimates can be highly sensitive to measurement noise. Regularization can improve stability but may introduce bias. Moreover, conventional NNLS-based methods provide point estimates without directly quantifying voxelwise estimation variability. To address these limitations, we propose the non-negative least squares Bootstrap Barycenter (NNLS-BB), which combines residual resampling with repeated NNLS reconstruction. We average spectra and derived parameters across bootstrap replicates and report parameter standard deviations to characterize sensitivity to signal perturbations. The method requires neither an explicit spectral regularization penalty nor a training dataset. NNLS-BB was evaluated using 40,000 simulated voxels with signal-to-noise ratios ranging from 50 to 1000, against conventional NNLS, two regularized NNLS methods, and a model-informed machine-learning (MIML) method. Scan–rescan repeatability was assessed across 20 white-matter regions in two independent cohorts comprising 13 healthy participants. NNLS-BB achieved the lowest MWF mean absolute error and root mean squared error in every simulated noise regime. Mean absolute error decreased by 36–41\% relative to conventional NNLS, alongside lower error dispersion. Compared with Bayes-Reg-\(D^{-1}\) and MIML, NNLS-BB achieved comparable dispersion but substantially lower bias, although MIML reconstructed the full spectrum more accurately. In vivo, NNLS-BB yielded the lowest regional scan–rescan dispersion and highest mean correlation in both cohorts. Bootstrap standard-deviation maps provided complementary information about parameter variability. These results support residual-bootstrap aggregation as a practical approach to improving MWF estimation while providing voxelwise variability maps from the same acquisition.
\end{abstract}

%%%%%%%%%%%%%%%%%%%%%%%%%%%%%%%%%%%%%%%%%%%%%%%%%%%%%%%%%%%%%%%%%%%%
\section{Introduction}\label{Introduction}
%%%%%%%%%%%%%%%%%%%%%%%%%%%%%%%%%%%%%%%%%%%%%%%%%%%%%%%%%%%%%%%%%%%%

Multicomponent $T_2$ relaxometry probes brain tissue microstructure
by separating MRI signal contributions associated with water pools
exhibiting different transverse relaxation properties. In white
matter, one of its main applications is myelin water imaging (MWI), where
the rapidly decaying signal component, typically associated with
$T_2$ values of approximately 10--30~ms, is commonly attributed to
water confined between the myelin bilayers. The relative contribution
of this short-$T_2$ component defines the myelin water fraction (MWF),
a widely studied MRI marker associated with myelin content.
MWF has been investigated in neurological disorders affecting
white-matter microstructure
\citep{Does2018,Alonso-Ortiz2015,Piredda2020ProbingReview,
Rivas-Fernandez2024BrainStudy},
as well as in normal aging
\citep{Canales-Rodriguez2021Age-Relaxometry}.

A conventional approach models the measured signal as a superposition of exponential relaxation components characterized by different $T_2$ values. Their amplitudes define a non-negative $T_2$ spectrum, which is typically estimated on a predefined $T_2$ grid using non-negative least squares (NNLS) \citep{Mackay1994,Whittall1997,MacKay2006}.
Although conceptually simple, this inverse problem is intrinsically
ill-conditioned because components with nearby $T_2$ values generate
very similar signal decays over the finite set of measured echo
times. Consequently, small perturbations of the data can produce
substantial changes in the recovered spectrum. Related studies
of multiexponential fitting, in which both component amplitudes
and relaxation times are estimated, have also shown that the identification of 
distinct components strongly depends on the signal-to-noise ratio
(SNR) and temporal sampling
\citep{Bertero1982OnNoise,Istratov1999ExponentialPhenomena}.

This instability is particularly problematic for myelin water
because the short-$T_2$ component decays rapidly and represents
only a small fraction of the total signal. MRI studies have shown
that reliable separation of relaxation components requires
a sufficiently high SNR
\citep{Graham1996CriteriaData,Andrews2005TestingRelaxometry},
while numerical and in vivo studies have shown that the
short-$T_2$ component can be missed under realistic acquisition
conditions, resulting in zero-valued MWF estimates even in
anatomically myelinated white-matter regions
\citep{Kumar2012BayesianMRI,Raj2014Multi-compartmentModel,
Canales-Rodriguez2021ComparisonQuantification}.
Together, these observations indicate that the MWF returned by
an NNLS reconstruction can strongly depend on the particular
noise realization.

Regularized NNLS (RNNLS) is commonly used to reduce this sensitivity.
By introducing an additional penalty, regularization stabilizes
the recovered spectrum and generally reduces the variability
of the derived MWF estimates. Common penalties favor spectra with
smaller norms or smoother profiles
\citep{Whittall1989,Graham1996CriteriaData,
Canales-Rodriguez2021ComparisonQuantification}.
Our previous studies revealed a particularly relevant trade-off
for MWF estimation: unregularized NNLS exhibited very small bias
but comparatively high variability, whereas RNNLS reduced
variability and overall MWF estimation error at the cost of
introducing a systematic bias
\citep{Canales-Rodriguez2021ComparisonQuantification,
Canales-Rodriguez2021RevisitingSquares}.

A second limitation is that conventional NNLS and RNNLS provide
a single reconstruction without directly quantifying the
sensitivity of the estimated spectrum and derived metrics
to measurement noise in each voxel. Characterizing this
variability would provide complementary information about
the reliability of the estimates.

Based on these considerations, we introduce the NNLS \emph{Bootstrap Barycenter} (NNLS-BB), which combines residual bootstrap resampling with repeated unregularized NNLS reconstruction. We summarize the bootstrap spectra using their Euclidean barycenter, which corresponds to their arithmetic mean, and average derived quantities such as MWF across the corresponding bootstrap estimates.
The dispersion of these estimates provides an empirical measure of noise-related variability, allowing voxelwise mapping of this variability for the spectrum-derived metrics.
Our central hypothesis is that aggregating these reconstructions can reduce noise-induced variability and MWF estimation error while preserving the low bias observed for unregularized NNLS in our previous studies.

We tested this hypothesis using numerical simulations and two independent in vivo scan--rescan datasets. In simulations, NNLS-BB is compared with conventional NNLS, two state-of-the-art RNNLS strategies, and a model-informed machine-learning method, assessing both the MWF estimation and the reconstruction of the full $T_2$ spectrum
under different noise conditions. In vivo, we evaluated
scan--rescan repeatability across white-matter regions and
examined voxelwise bootstrap-variability maps for several
spectrum-derived metrics. 

The proposed method has been incorporated into our multicomponent $T_2$ reconstruction toolbox, available at \url{https://github.com/ejcanalesr/multicomponent-T2-toolbox}.

%%%%%%%%%%%%%%%%%%%%%%%%%%%%%%%%%%%%%%%%%%%%%%%%%%%%%%%%%%%%%%%%%%%%
\section{Theory}
\label{sec:theory}
%%%%%%%%%%%%%%%%%%%%%%%%%%%%%%%%%%%%%%%%%%%%%%%%%%%%%%%%%%%%%%%%%%%%

\subsection{Signal model and NNLS estimation}

The measured multi-echo signal is modeled as
\begin{equation}
\mathbf{y}
=
\mathbf{A}\mathbf{x}_0+\boldsymbol{\varepsilon},
\qquad
\boldsymbol{\varepsilon}
\sim\mathcal{N}(\mathbf{0},\sigma^2\mathbf{I}),
\qquad
\mathbf{x}_0\geq\mathbf{0},
\label{eq:linear_inverse_problem}
\end{equation}
where $\mathbf{y}\in\mathbb{R}^{N}$ contains the measurements
at $N$ echo times, $\mathbf{x}_0\in\mathbb{R}^{P}$ is the
underlying spectrum on a grid of $P$ relaxation times,
and $\sigma$ is the noise standard deviation.
Each column of $\mathbf{A}\in\mathbb{R}^{N\times P}$
describes the signal associated with one point of the grid,
using either a mono-exponential decay model or the
Extended Phase Graph formalism (EPG)
\citep{Hennig1988MultiechoAngles,Weigel2015,
Prasloski2012a,Prasloski2012}.

Under this Gaussian noise model, maximizing the likelihood
over non-negative spectra is equivalent to minimizing
the squared residual:
\begin{equation}
\widehat{\mathbf{x}}_{\mathrm{nnls}}
\in
\operatorname*{arg\,min}_{\mathbf{x}\geq\mathbf{0}}
\|\mathbf{y}-\mathbf{A}\mathbf{x}\|_2^2,
\label{eq:nnls_ML}
\end{equation}
where $\widehat{\mathbf{x}}_{\mathrm{nnls}}$ denotes the
solution returned by the NNLS fitting procedure.

\subsection{Spectral ambiguity and the effect of noise}

The sensitivity of linear inverse problems to noise is
closely related to the singular values of the forward
model \citep{Hansen2010DiscreteProblems}.
In multiexponential reconstruction, highly correlated dictionary columns allow substantially different spectra to produce very similar signals. Yet these spectra may assign different fractions of their total amplitude to the short-\(T_2\) range, leading to different MWF estimates despite similarly good fits to the measured signal. The geometry of the fitting objective helps explain how such ambiguities arise.

Let $\mathbf{v}_k$ be a unit eigenvector of
$\mathbf{G}=\mathbf{A}^{\top}\mathbf{A}$ with eigenvalue
$g_k\geq0$.
For a displacement from the true spectrum along
$\mathbf{v}_k$, the squared difference from the
noise-free signal is
\begin{equation}
\begin{split}
\Phi_0(\mathbf{x}_0+t\mathbf{v}_k)
&=
\|\mathbf{A}(\mathbf{x}_0+t\mathbf{v}_k)
-\mathbf{A}\mathbf{x}_0\|_2^2
\\
&=
t^2\|\mathbf{A}\mathbf{v}_k\|_2^2
=
g_k t^2.
\end{split}
\label{eq:signal_variation_eigen}
\end{equation}

Thus, when \(g_k\) is small, relatively large changes in the spectrum along \(\mathbf{v}_k\) produce only small changes in the predicted signal. Such displacements are admissible only if \(\mathbf{x}_0+t\mathbf{v}_k\geq\mathbf{0}\), but within this constraint they are weakly distinguished by the measurements. We next examine how measurement noise can influence which of these nearly equivalent spectra is favored by NNLS.

For the measured signal, define the fitting cost function
$\Phi_{\mathbf{y}}(\mathbf{x})
=\|\mathbf{y}-\mathbf{A}\mathbf{x}\|_2^2$.
Its global minimum over non-negative spectra is attained
by $\widehat{\mathbf{x}}_{\mathrm{nnls}}$.
We use the true spectrum $\mathbf{x}_0$ as a reference
for studying the effect of noise.
Substituting
$\mathbf{y}=\mathbf{A}\mathbf{x}_0+\boldsymbol{\varepsilon}$
gives the relationship
\begin{equation}
\begin{aligned}
\Phi_{\mathbf{y}}(\mathbf{x}_0+t\mathbf{v}_k)
-\Phi_{\mathbf{y}}(\mathbf{x}_0)
&=
\|\boldsymbol{\varepsilon}-t\mathbf{A}\mathbf{v}_k\|_2^2
-\|\boldsymbol{\varepsilon}\|_2^2 \\
&= g_k t^2-2tz_k \\
&= g_k\left(t-\frac{z_k}{g_k}\right)^2
-\frac{z_k^2}{g_k},
\qquad g_k>0,
\end{aligned}
\label{eq:noise_displacement}
\end{equation}
where $z_k=(\mathbf{A}\mathbf{v}_k)^\top
\boldsymbol{\varepsilon}$.

The last expression describes a parabola in $t$.
Its second derivative, $2g_k$, characterizes how
sharply the fitting cost increases away from its
minimum and is independent of noise.
The noise-dependent term $z_k$, instead, shifts
the position of this minimum.
In the absence of noise, the minimum occurs at
$t=0$, corresponding to the true spectrum
$\mathbf{x}_0$.
For $g_k>0$, noise shifts the unconstrained minimum
to $t=z_k/g_k$.
Along directions with small $g_k$, this displacement
can be large, provided it remains compatible with
the non-negativity constraint.

Whenever an admissible displacement satisfies
$2tz_k>g_k t^2$, the displaced spectrum fits the
noisy measurement more closely than the true
spectrum does.
Setting $g_k t^2-2tz_k=0$ identifies spectra
with the same fitting cost as $\mathbf{x}_0$.
If $t=2z_k/g_k$ is well-defined and non-zero,
and the displaced spectrum remains non-negative,
the true and displaced spectra lie on the same
level surface of the cost function.

Together, directions associated with small
eigenvalues give rise to a shallow valley of
similar fitting errors, where different spectra
can explain the measured signal almost equally well.
The level surface passing through $\mathbf{x}_0$
has cost
$\Phi_{\mathbf{y}}(\mathbf{x}_0)
=\|\boldsymbol{\varepsilon}\|_2^2$,
and the NNLS solution lies at or below this level.
Across repeated acquisitions, noise can change
the location of the minimum, potentially altering
both the selected $T_2$ bins and derived parameters
such as MWF.

Regularized NNLS addresses this sensitivity by introducing
a preference for particular spectral properties:
\begin{equation}
\widehat{\mathbf{x}}_{\mathrm{rnnls}}
\in
\operatorname*{arg\,min}_{\mathbf{x}\geq\mathbf{0}}
\left\{
\|\mathbf{y}-\mathbf{A}\mathbf{x}\|_2^2
+
\lambda\|\mathbf{L}\mathbf{x}\|_2^2
\right\},
\label{eq:map}
\end{equation}
where $\lambda\geq0$ controls the penalty strength.
The penalty increases the coefficient of the quadratic
term along $\mathbf{v}_k$ from $g_k$ to
$g_k+\lambda\|\mathbf{L}\mathbf{v}_k\|_2^2$.
Geometrically, this makes the regularized objective steeper along directions affected by the penalty, narrowing the range of spectral displacements with similarly low objective values. Regularization can therefore reduce the sensitivity of the reconstruction to noise-driven fluctuations.
For zero-order regularization, $\mathbf{L}=\mathbf{I}$,
this increase is constant and equal to $\lambda$ in every direction, and
the penalty favors spectra with smaller norms.
First- and second-order regularization penalize first
and second differences between spectral amplitudes,
respectively, promoting smoother estimates.

However, the structural preferences promoted by regularization can also bias derived parameters through attenuation, broadening, or redistribution of spectral components \citep{Canales-Rodriguez2021ComparisonQuantification,Canales-Rodriguez2021RevisitingSquares}. Consequently, a penalty suitable for reconstructing the full spectrum need not be optimal for estimating MWF.

\subsection{Aggregation of perturbed reconstructions}

As discussed in the previous section, substantially different spectra can explain the measured signal almost equally well when the fitting objective has a relatively flat valley. The data alone then provide little basis for reliably preferring one spectrum over another, making the choice of a single representative spectrum ambiguous. Rather than imposing additional spectral preferences through regularization, an alternative is to combine NNLS estimates obtained under different noise perturbations, allowing their fluctuations to partially compensate through averaging. In principle, repeated acquisitions of the same underlying signal could provide such an ensemble, but this would be experimentally costly. In this work, NNLS-BB therefore uses a residual bootstrap to approximate this ensemble from a single acquisition.

Let $\mathbf{x}_1,\ldots,\mathbf{x}_B$ denote an ensemble
of non-negative reconstructions.
Their Euclidean barycenter is their arithmetic mean,
\begin{equation}
\overline{\mathbf{x}}
=
\frac{1}{B}\sum_{b=1}^{B}\mathbf{x}_b,
\label{eq:theory_ensemble_mean}
\end{equation}
which remains non-negative.
For a parameter $\theta(\mathbf{x})$ derived from the
spectrum, define $\theta_b=\theta(\mathbf{x}_b)$ and the mean value
$\overline{\theta}=B^{-1}\sum_b\theta_b$.
The effect of averaging under squared error is described
by the classical ambiguity decomposition
\citep{Krogh1994NeuralLearning}:
\begin{equation}
(\overline{\theta}-\theta_0)^2
=
\frac{1}{B}\sum_{b=1}^{B}(\theta_b-\theta_0)^2
-
\frac{1}{B}\sum_{b=1}^{B}
(\theta_b-\overline{\theta})^2,
\label{eq:ensemble_parameter_error}
\end{equation}
where $\theta_0$ is the true parameter value.

The three terms in this identity have a direct interpretation: the left-hand side is the squared error of the ensemble mean, the first term on the right-hand side is the average squared error of the individual estimates, and the second is their empirical variance. Since this variance is non-negative, the ensemble mean always achieves a squared error no greater than the average squared error of its members, with a strict improvement whenever the individual estimates differ. Thus, although averaging does not necessarily outperform the best individual estimate, which is generally unknown, it guarantees a lower squared error than the ensemble members achieve on average. This property is particularly appealing for ill-conditioned inverse problems, where noise perturbations can produce substantially different estimates whose fluctuations may compensate through aggregation.

For nonlinear parameters such as MWF, \(\overline{\theta}\) generally differs from \(\theta(\overline{\mathbf{x}})\). We therefore average the derived parameter values directly, alongside the spectrum-level barycenter.

%%%%%%%%%%%%%%%%%%%%%%%%%%%%%%%%%%%%%%%%%%%%%%%%%%%%%%%%%%%%%%%%%%%%
\section{Methods}
%%%%%%%%%%%%%%%%%%%%%%%%%%%%%%%%%%%%%%%%%%%%%%%%%%%%%%%%%%%%%%%%%%%%

We implement NNLS-BB by generating perturbed signals
through residual resampling and fitting each signal
with NNLS. The resulting spectra and derived parameter
values are averaged after MWF-based trimming,
and their variability is summarized using bootstrap
standard deviations.

\subsection{Residual-bootstrap reconstructions}

For each measured signal $\mathbf{y}$, we first compute
$\widehat{\mathbf{x}}_{\mathrm{nnls}}$ using
Eq.~\eqref{eq:nnls_ML}.
NNLS problems were solved using the active-set algorithm
\citep{Lawson1974}.
The fitted signal is
$\widehat{\mathbf{y}}
=\mathbf{A}\widehat{\mathbf{x}}_{\mathrm{nnls}}$,
and the raw residuals are
$\mathbf{r}=\mathbf{y}-\widehat{\mathbf{y}}$.

Let $\mathcal{P}$ denote the active set of the NNLS
estimate, i.e., the indices of its non-zero coefficients,
and let $\mathbf{A}_{\mathcal{P}}$ contain the corresponding
columns of $\mathbf{A}$.
Using the active set as a local linear approximation to the fit, we first mean-center the residuals and then apply a standard leverage adjustment to compensate for differences in their marginal variances before bootstrap resampling \citep{Davison1997BootstrapApplication}:
\begin{equation}
r_i'
=
\frac{r_i-\bar r}{\sqrt{1-h_{ii}}},
\label{eq:adjusted_residuals}
\end{equation}
where $\bar r$ is the empirical mean of the raw residuals
and $h_{ii}$ is the $i$th diagonal element of the hat matrix
\[
\mathbf{H}
=
\mathbf{A}_{\mathcal{P}}
\left(
\mathbf{A}_{\mathcal{P}}^{\top}\mathbf{A}_{\mathcal{P}}
\right)^{\dagger}
\mathbf{A}_{\mathcal{P}}^{\top}.
\]
Here, $\dagger$ denotes the Moore--Penrose pseudoinverse.

For each bootstrap replicate $b=1,\ldots,B$, we construct
a residual vector $\mathbf{r}^{*(b)}$ by independently
sampling $N$ entries with replacement from
$\{r_i'\}_{i=1}^{N}$.
We then form the perturbed signal
\begin{equation}
\mathbf{y}^{*(b)}
=
\widehat{\mathbf{y}}+\mathbf{r}^{*(b)}
\label{eq:bootstrap_replicate}
\end{equation}
and reconstruct it using the same NNLS procedure,
obtaining $\widehat{\mathbf{x}}^{*(b)}$.

\subsubsection{Trimmed bootstrap barycenters and variability}

To limit the influence of extreme bootstrap estimates,
we applied trimming at both tails of the MWF ensemble.
Let $q_{\alpha}$ and $q_{1-\alpha}$ denote the empirical
quantiles of the bootstrap MWF values.
The retained indices and their number are
\begin{equation}
\begin{split}
\mathcal{I}_{\alpha}
&=
\left\{
b:
q_{\alpha}
\leq
\theta_{\mathrm{MWF}}
\!\left(\widehat{\mathbf{x}}^{*(b)}\right)
\leq
q_{1-\alpha}
\right\},
\\
K&=|\mathcal{I}_{\alpha}|.
\end{split}
\label{eq:trimmed_index_set}
\end{equation}
We fixed $\alpha=0.005$, nominally excluding the most
extreme $1\%$ of bootstrap estimates.
Values equal to either quantile were retained.

The same retained replicates were used to average
the spectra and all derived parameters:
\begin{equation}
\begin{split}
\widehat{\mathbf{x}}_{\mathrm{BB}}
&=
\frac{1}{K}
\sum_{b\in\mathcal{I}_{\alpha}}
\widehat{\mathbf{x}}^{*(b)},
\\
\widehat{\theta}_{\mathrm{BB}}
&=
\frac{1}{K}
\sum_{b\in\mathcal{I}_{\alpha}}
\theta\!\left(\widehat{\mathbf{x}}^{*(b)}\right).
\end{split}
\label{eq:trimmed_barycenters}
\end{equation}
The spread of each parameter ensemble was
summarized using its bootstrap standard deviation
\citep{Efron1986BootstrapAccuracy}:
\begin{equation}
\operatorname{SD}_{\mathrm{boot}}(\theta)
=
\sqrt{
\frac{1}{K-1}
\sum_{b\in\mathcal{I}_{\alpha}}
\left[
\theta\!\left(\widehat{\mathbf{x}}^{*(b)}\right)
-\widehat{\theta}_{\mathrm{BB}}
\right]^2
}.
\label{eq:trimmed_bootstrap_standard_deviation}
\end{equation}
All reported NNLS-BB estimates and
bootstrap-variability summaries were computed following this procedure.

\subsection{Metrics derived from the $T_2$ spectrum}

We estimated MWF, total water content (TWC),
intra- and extracellular water fraction (IEWF),
quasi-free water fraction (QFWF), and the geometric
mean $T_2$ values of the myelin-water and intra- and
extracellular-water compartments, denoted by
$T_2^{\mathrm{MW}}$ and $T_2^{\mathrm{IE}}$, respectively.

The spectrum was partitioned into three segments
using cutoff values commonly adopted in myelin water
imaging \citep{MacKay2006}.
The upper limits of the myelin-water and intra- and
extracellular-water ranges were set to
$T_2^{\mathrm{m}}=40~\mathrm{ms}$ and
$T_2^{\mathrm{i}}=200~\mathrm{ms}$, respectively.
The corresponding spectral-bin indices were
\begin{equation}
\begin{aligned}
\mathcal{I}_{\mathrm{MW}}
&=
\left\{j:T_{2,j}\leq T_2^{\mathrm{m}}\right\},
\\
\mathcal{I}_{\mathrm{IE}}
&=
\left\{j:T_2^{\mathrm{m}}<T_{2,j}
\leq T_2^{\mathrm{i}}\right\},
\\
\mathcal{I}_{\mathrm{QF}}
&=
\left\{j:T_{2,j}>T_2^{\mathrm{i}}\right\}.
\end{aligned}
\label{eq:mw_indices}
\end{equation}
TWC was estimated from the sum of the spectral amplitudes \citep{Meyers2017}
and is reported in arbitrary signal units:
\begin{equation}
\theta_{\mathrm{TWC}}(\mathbf{x})
=
\sum_{j=1}^{P}x_j.
\label{eq:twc_definition}
\end{equation}
The compartment fractions were computed by dividing
their respective amplitudes by the total:
\begin{equation}
\begin{aligned}
\theta_{\mathrm{MWF}}(\mathbf{x})
&=
\frac{\sum_{j\in\mathcal{I}_{\mathrm{MW}}}x_j}
{\sum_{j=1}^{P}x_j},
\\
\theta_{\mathrm{IEWF}}(\mathbf{x})
&=
\frac{\sum_{j\in\mathcal{I}_{\mathrm{IE}}}x_j}
{\sum_{j=1}^{P}x_j},
\\
\theta_{\mathrm{QFWF}}(\mathbf{x})
&=
\frac{\sum_{j\in\mathcal{I}_{\mathrm{QF}}}x_j}
{\sum_{j=1}^{P}x_j}.
\end{aligned}
\label{eq:compartment_fractions}
\end{equation}
The geometric mean $T_2$ values were calculated
as amplitude-weighted means on the logarithmic scale
\citep{Bjarnason2011ProofGmR2}:
\begin{equation}
\begin{aligned}
\theta_{T_2^{\mathrm{MW}}}(\mathbf{x})
&=
\exp\left(
\frac{
\sum_{j\in\mathcal{I}_{\mathrm{MW}}}
x_j\log(T_{2,j})
}{
\sum_{j\in\mathcal{I}_{\mathrm{MW}}}x_j
}
\right),
\\
\theta_{T_2^{\mathrm{IE}}}(\mathbf{x})
&=
\exp\left(
\frac{
\sum_{j\in\mathcal{I}_{\mathrm{IE}}}
x_j\log(T_{2,j})
}{
\sum_{j\in\mathcal{I}_{\mathrm{IE}}}x_j
}
\right).
\end{aligned}
\label{eq:geometric_mean_t2}
\end{equation}

\subsection{Evaluated algorithms}

We compared NNLS-BB with four reference methods. All five methods were evaluated in the numerical simulations, whereas the in vivo comparison included only the four NNLS-based methods.

The NNLS-based reference methods included standard NNLS, computed using the SciPy implementation (SciPy NNLS). We also evaluated conventional RNNLS with identity regularization, \(\mathbf{L}=\mathbf{I}\), and a \(\chi^2\)-based criterion for selecting the regularization weight (NNLS-\(\chi^2\)-I). The weight was chosen so that the sum of squared residuals was 2\% higher than that of the unregularized NNLS solution \citep{Whittall1989,Graham1996CriteriaData,Laule2006,MacKay2006}. 
Finally, we included Bayesian regularized NNLS with the inverse-distance regularization matrix \(\mathbf{L}=\mathbf{D}^{-1}\), referred to as Bayes-Reg-\(D^{-1}\) \citep{Canales-Rodriguez2021RevisitingSquares}.

As a supervised-learning benchmark, we also included Model-Informed Machine Learning (MIML), a method for \(T_2\) spectrum estimation \citep{Yu2020Model-InformedRelaxometry}. MIML uses a multilayer perceptron trained to map multi-echo signals to their corresponding ground-truth \(T_2\) distributions. We retrained the model using the same network architecture, synthetic training \(T_2\) spectra, signal model, and acquisition parameters described in \citep{Yu2020Model-InformedRelaxometry}. The training SNR range was extended from 80–200 to 50–500 to provide broader coverage of the noise conditions evaluated in this study. We also increased the maximum number of training epochs from 30 to 100 and used early stopping to limit overfitting.

\subsection{Experimental data and evaluation protocol}

\subsubsection{Numerical simulations and evaluation metrics}

The algorithms were first evaluated using synthetic
multi-echo $T_2$ data with known ground-truth spectra
and MWF values. The simulated spectra consisted of
two Gaussian components representing myelin water
and the combined intra- and extracellular water pool.
Component amplitudes, mean $T_2$ values, spectral widths,
refocusing flip angles, and SNR values were randomly
sampled to cover a broad range of white-matter-like
signal configurations. The simulation parameters
are reported in Table~\ref{tab:simulation_parameters}.

Signal generation followed the implementation described
in \citep{Canales-Rodriguez2021RevisitingSquares}.
To reduce discretization error, each Gaussian spectrum
was represented on a dense $T_2$ grid containing
1000 equally spaced points between 1 and 300~ms.
The multi-echo signal was computed as the weighted
sum of the EPG signals associated with these points,
using the spectral amplitudes as weights.
For reconstruction and spectral comparison,
the ground-truth spectra were represented on the
60-point $T_2$ grid shared by all evaluated methods \citep{Canales-Rodriguez2021RevisitingSquares}.

The simulated signals were contaminated with Rician
noise. SNR was defined as the noise-free signal amplitude
at the shortest echo time divided by the standard
deviation of the underlying Gaussian noise.
Four independent datasets were generated by uniformly
sampling SNR within the intervals $50$--$100$,
$100$--$200$, $200$--$400$, and $400$--$1000$.
Each dataset contained $10{,}000$ independent voxels,
yielding $40{,}000$ simulated voxels in total.

The evaluation considered both MWF estimation and
reconstruction of the full $T_2$ spectrum.
MWF accuracy was summarized using mean absolute error
(MAE), root mean squared error (RMSE), centered RMSE
(cRMSE), mean bias error (MBE), and Pearson's
correlation coefficient $R$.
The decomposition
$\mathrm{RMSE}^2=\mathrm{cRMSE}^2+\mathrm{MBE}^2$
separates the centered estimation error from the
overall systematic bias
\citep{Taylor2001SummarizingDiagram}.
The reconstructed and ground-truth spectra were
normalized and compared using spectral RMSE
(RMSE-S) and mean Jensen--Shannon distance
(MJSD-S) \citep{Endres2003ADistributions}.
Computational runtime was also recorded for each
method and SNR regime.

To assess the consistency of the MWF error differences
across groups of simulated voxels, each dataset was
divided into 50 non-overlapping blocks of 200 voxels.
For each block and method, we computed the mean
absolute MWF error. NNLS-BB was compared with each
reference method using the Wilcoxon signed-rank test,
with the errors from the same blocks forming paired
observations. Holm correction was applied separately
within each SNR regime to account for the four
comparisons. We also recorded the percentage of blocks
in which NNLS-BB achieved a lower MAE than each
reference method.

The effect of the number of bootstrap replicates
was evaluated over ensemble sizes up to $B=1000$,
using standard NNLS as the reference without aggregation.
For each SNR regime, we examined the MWF MAE and its
ratio to the MAE obtained with $B=1000$.
This analysis informed the choice of $B=200$
for the remaining evaluations.

\begin{table}[!t]
\centering
\caption{Parameters used to generate the synthetic white-matter \(T_2\)-relaxometry datasets.}
\label{tab:simulation_parameters}
\scriptsize
\setlength{\tabcolsep}{4.5pt}
\renewcommand{\arraystretch}{1.15}
\begin{tabular}{p{0.47\columnwidth}p{0.43\columnwidth}}
\toprule
Parameter & Range or value \\
\midrule
Flip angle & \(\alpha \sim \mathcal{U}(90^\circ,180^\circ)\) \\
SNR regimes & \(50{-}100,\ 100{-}200,\)
              \newline \(200{-}400,\ 400{-}1000\) \\\midrule
Myelin water fraction & \(\mathrm{MWF} \sim \mathcal{U}(0,0.25)\) \\
Intra- and extracellular water fraction & \(\mathrm{IEWF}=1-\mathrm{MWF}\) \\
\midrule
Myelin \(T_2\) & \(T_{2,\mathrm{m}} \sim \mathcal{U}(15,35)~\mathrm{ms}\) \\
Intra- and extracellular \(T_2\) & \(T_{2,\mathrm{IE}} \sim \mathcal{U}(60,90)~\mathrm{ms}\) \\
\midrule
Myelin lobe width & \(\sigma_{\mathrm{m}} \sim \mathcal{U}(1,3)~\mathrm{ms}\) \\
Intra- and extracellular lobe width & \(\sigma_{\mathrm{IE}} \sim \mathcal{U}(6,12)~\mathrm{ms}\) \\
\bottomrule
\end{tabular}

\vspace{1mm}
\footnotesize
\(\mathcal{U}(a,b)\) denotes a uniform distribution on \([a,b]\).
\end{table}

\subsubsection{Human brain data and repeatability metrics}

Scan--rescan data from two independent cohorts were used
to evaluate the repeatability of MWF estimates.
Each participant underwent two acquisitions.

Cohort~1 initially included eight healthy volunteers
aged 23--33 years, scanned in two consecutive sessions
on a 3T scanner
(MAGNETOM Prisma, Siemens Healthcare, Forchheim, Germany) using a
64-channel head/neck coil.
Three participants were excluded because of motion
artifacts, leaving five paired datasets for analysis.
Data were acquired using a high-resolution 3D
gradient and spin-echo (GRASE) research application sequence accelerated
with CAIPIRINHA \citep{Piredda2021FastCAIPIRINHA}.
The acquisition parameters were:
matrix size $144\times126\times134$,
voxel size $1.6\times1.6\times1.6~\mathrm{mm}^3$,
32 echoes with an initial TE and echo spacing
of 10.68~ms, TR $=1~\mathrm{s}$,
prescribed refocusing flip angle $=180^\circ$,
acceleration factor $3\times2$, and one average.
The acquisition time was 10~min 30~s.
The study was approved by the local Institutional
Review Board, and written informed consent was
obtained before each examination.

Cohort~2 initially included ten healthy volunteers (age range: 20--25 years), scanned on a Philips 3T Achieva scanner
(Philips Medical Systems, Best, The Netherlands).
Two participants were excluded because of motion
artifacts, leaving eight paired datasets for analysis.
Multi-echo data were acquired using a 3D GRASE
sequence with the protocol described in
\citep{Rivas-Fernandez2024BrainStudy}.
The acquisition used 32 echoes, an echo spacing
of 10.24~ms, TR $=2~\mathrm{s}$, and a voxel size
of $1.7\times1.7\times4~\mathrm{mm}^3$.
The study was approved by the Galician Clinical
Research Ethics Committee
(Xunta de Galicia, Spain; references 2017/498
and 2022/116) and conducted in accordance with
the Declaration of Helsinki.
All participants provided written informed consent.

Scan--rescan repeatability was evaluated using
regional MWF values from the two acquisitions.
Regions were defined using the 20 white-matter
tracts of the JHU white-matter tractography atlas \citep{Hua2007TractQuantification}
distributed with FSL \citep{Jenkinson2011}.
We used the maximum-probability label atlas
at 1~mm resolution.

For each subject and acquisition, a registration
transform to the common atlas space was estimated
using ANTs \citep{Avants2008}.
The 10th echo image was used for registration,
corresponding to TE $=106.8~\mathrm{ms}$ in Cohort~1
and TE $=102.4~\mathrm{ms}$ in Cohort~2.
The same subject- and acquisition-specific transform
was applied to the MWF maps from all reconstruction
methods, ensuring a common anatomical basis for
their comparison.

For each subject $s$ and region of interest (ROI) $i$, the scan--rescan difference was defined as $d_{s,i} = x_{s,i}^{(2)}-x_{s,i}^{(1)}$, where $x_{s,i}^{(1)}$ and $x_{s,i}^{(2)}$ are the mean MWF
values in the scan and rescan acquisitions, respectively.

The mean scan--rescan offset and the standard
deviation of the regional differences were calculated
separately for each subject:
\begin{align}
\mathrm{MBE}_s
&=
\frac{1}{R}\sum_{i=1}^{R}d_{s,i},
\label{eq:subject_mbe}
\\
\mathrm{SD}_{\mathrm{diff},s}
&=
\sqrt{
\frac{1}{R-1}
\sum_{i=1}^{R}
\left(d_{s,i}-\mathrm{MBE}_s\right)^2
},
\label{eq:subject_sddiff}
\end{align}
where $R=20$ is the number of white-matter tract ROIs. We also calculated Pearson's correlation coefficient
$r_s$ and the slope $\beta_s$ of a least-squares regression of rescan against scan
MWF values across ROIs, including an intercept.

The subject-specific metrics were summarized
separately for each cohort.
Mean $|\mathrm{MBE}|$ was calculated by averaging
the absolute subject-specific offsets, and mean
$\mathrm{SD}_{\mathrm{diff}}$ by averaging the
subject-specific standard deviations.
These measures describe the magnitude of the
scan--rescan offset and the regional dispersion
of the differences, respectively.
The correlation coefficients were summarized
by their mean and sample standard deviation
(mean $r$ and SD $r$).
Finally, mean $|\beta-1|$ was calculated as the
average absolute deviation of the subject-specific
regression slopes from unity.

Bland--Altman analysis was performed for NNLS-BB
by pooling the regional scan--rescan differences
across subjects within each cohort. For qualitative assessment, we generated axial scan--rescan MWF maps. For NNLS-BB, each spectrum-derived parameter map was accompanied by a map of the standard deviation of its bootstrap estimates.

\subsection{Preprocessing and estimation}

The human brain multi-echo images were denoised before
spectral fitting using local principal component analysis
\citep{Manjon2013}, implemented in DIPY
\citep{Garyfallidis2014}.
Denoising was applied jointly across echoes using
$5\times5\times5$-voxel spatial patches.
The noise standard deviation map was obtained using
\texttt{pca\_noise\_estimate}, with
\texttt{correct\_bias=False} and \texttt{smooth=3}.
We used \texttt{tau\_factor=1}, setting the local
PCA eigenvalue threshold to the estimated noise
variance.
This lower threshold was chosen empirically to
retain more components and limit excessive smoothing
of the multi-echo data.

For all NNLS-based reconstructions, the forward-model
matrices were generated using the EPG formalism
\citep{Prasloski2012a}.
The $T_2$ spectrum was represented on a logarithmically
spaced grid of $P=60$ points spanning
10--2000~ms \citep{Prasloski2012}.
To account for differences between the prescribed
and effective refocusing flip angles caused by
$B_1$ inhomogeneity, a voxelwise flip-angle estimate
was obtained before spectral inversion, following
the procedure described in
\citep{Canales-Rodriguez2021ComparisonQuantification,
Canales-Rodriguez2021RevisitingSquares}.

To reduce computation time, EPG dictionaries were
precomputed for refocusing flip angles from
$90^\circ$ to $180^\circ$ at approximately
$0.33^\circ$ intervals.
Each voxel was assigned the dictionary corresponding
to the grid point closest to its estimated flip angle.
The selected dictionary was used for spectral
estimation with each NNLS-based reconstruction method.

\section{Results}

\subsection{Numerical simulations}

\begin{table*}[t]
\centering
\caption{
Quantitative comparison of the evaluated methods using synthetic magnitude data across SNR ranges. For each SNR regime, \(10{,}000\) simulated voxels were evaluated. MAE, RMSE, cRMSE, MBE, and \(R\) were computed for the myelin water fraction (MWF), whereas RMSE-S and MJSD-S were computed over the full normalized \(T_2\) spectrum. Runtime is reported in seconds for the complete evaluation set. Lower values indicate better performance for MAE, RMSE, cRMSE, RMSE-S, MJSD-S, and runtime; MBE values closer to zero and higher correlation coefficients \(R\) indicate better performance. Within each SNR regime and metric, the best value is shown in bold, and the second-best value is underlined.
}
\label{tab:multi_snr_metrics}
\resizebox{\textwidth}{!}{%
\begin{tabular}{llrrrrrrrr}
\toprule
SNR & Method & MAE & RMSE & cRMSE & MBE & R & RMSE-S & MJSD-S & Runtime (s) \\
\midrule
50-100 & SciPy NNLS & 0.067091 & 0.088333 & 0.085895 & \underline{-0.020613} & 0.534630 & 0.073830 & 0.579873 & \textbf{0.38} \\
 & NNLS-$\chi^2$-I & 0.054153 & 0.068936 & 0.062237 & -0.029644 & 0.682669 & 0.030734 & 0.380346 & 25.63 \\
 & Bayes-Reg-$D^{-1}$ & \underline{0.046413} & \underline{0.059033} & \underline{0.054951} & -0.021571 & \textbf{0.738118} & \underline{0.029808} & \underline{0.358482} & 19.29 \\
 & NNLS-BB & \textbf{0.043027} & \textbf{0.054874} & \textbf{0.054873} & \textbf{-0.000318} & 0.682236 & 0.039816 & 0.373827 & 48.28 \\
 & MIML & 0.048503 & 0.062754 & 0.056094 & -0.028134 & \underline{0.719693} & \textbf{0.019928} & \textbf{0.261199} & \underline{1.26} \\
\midrule
100-200 & SciPy NNLS & 0.051085 & 0.070493 & 0.069872 & \underline{-0.009341} & 0.675314 & 0.068797 & 0.553709 & \textbf{0.40} \\
 & NNLS-$\chi^2$-I & 0.042476 & 0.055593 & 0.045465 & -0.031994 & 0.805629 & 0.026601 & 0.336603 & 26.17 \\
 & Bayes-Reg-$D^{-1}$ & 0.038233 & 0.049357 & \underline{0.040949} & -0.027555 & \underline{0.846379} & \underline{0.025475} & \underline{0.312904} & 21.53 \\
 & NNLS-BB & \textbf{0.032009} & \textbf{0.043048} & 0.042484 & \textbf{-0.006947} & 0.818186 & 0.040360 & 0.360320 & 51.21 \\
 & MIML & \underline{0.037308} & \underline{0.049084} & \textbf{0.038764} & -0.030110 & \textbf{0.852871} & \textbf{0.016597} & \textbf{0.232323} & \underline{1.24} \\
\midrule
200-400 & SciPy NNLS & 0.043123 & 0.060602 & 0.060588 & \textbf{0.001291} & 0.759943 & 0.064458 & 0.532404 & \textbf{0.42} \\
 & NNLS-$\chi^2$-I & 0.034532 & 0.046175 & 0.037359 & -0.027137 & 0.866818 & 0.022628 & 0.293557 & 26.76 \\
 & Bayes-Reg-$D^{-1}$ & \underline{0.031738} & \underline{0.041916} & \underline{0.032948} & -0.025912 & \underline{0.896733} & \underline{0.021736} & \underline{0.276292} & 24.06 \\
 & NNLS-BB & \textbf{0.025376} & \textbf{0.034626} & 0.034553 & \underline{-0.002249} & 0.886884 & 0.039588 & 0.344063 & 56.18 \\
 & MIML & 0.034778 & 0.045199 & \textbf{0.031973} & -0.031947 & \textbf{0.897514} & \textbf{0.015569} & \textbf{0.223697} & \underline{1.23} \\
\midrule
400-1000 & SciPy NNLS & 0.037729 & 0.052581 & 0.051244 & \underline{0.011780} & 0.834520 & 0.061449 & 0.513243 & \textbf{0.45} \\
 & NNLS-$\chi^2$-I & 0.025973 & 0.036238 & 0.030401 & -0.019724 & 0.913791 & 0.018468 & 0.241933 & 27.32 \\
 & Bayes-Reg-$D^{-1}$ & \underline{0.023949} & \underline{0.033534} & \textbf{0.027946} & -0.018536 & \underline{0.927558} & \underline{0.017827} & \underline{0.234744} & 27.23 \\
 & NNLS-BB & \textbf{0.022560} & \textbf{0.030378} & \underline{0.029366} & \textbf{0.007777} & \textbf{0.928634} & 0.041790 & 0.344708 & 60.99 \\
 & MIML & 0.034407 & 0.044475 & 0.030192 & -0.032657 & 0.909523 & \textbf{0.015265} & \textbf{0.221365} & \underline{1.22} \\
\bottomrule
\end{tabular}%
}
\end{table*}

\begin{figure*}[!t]
  \centering
  \includegraphics[width=1\textwidth]{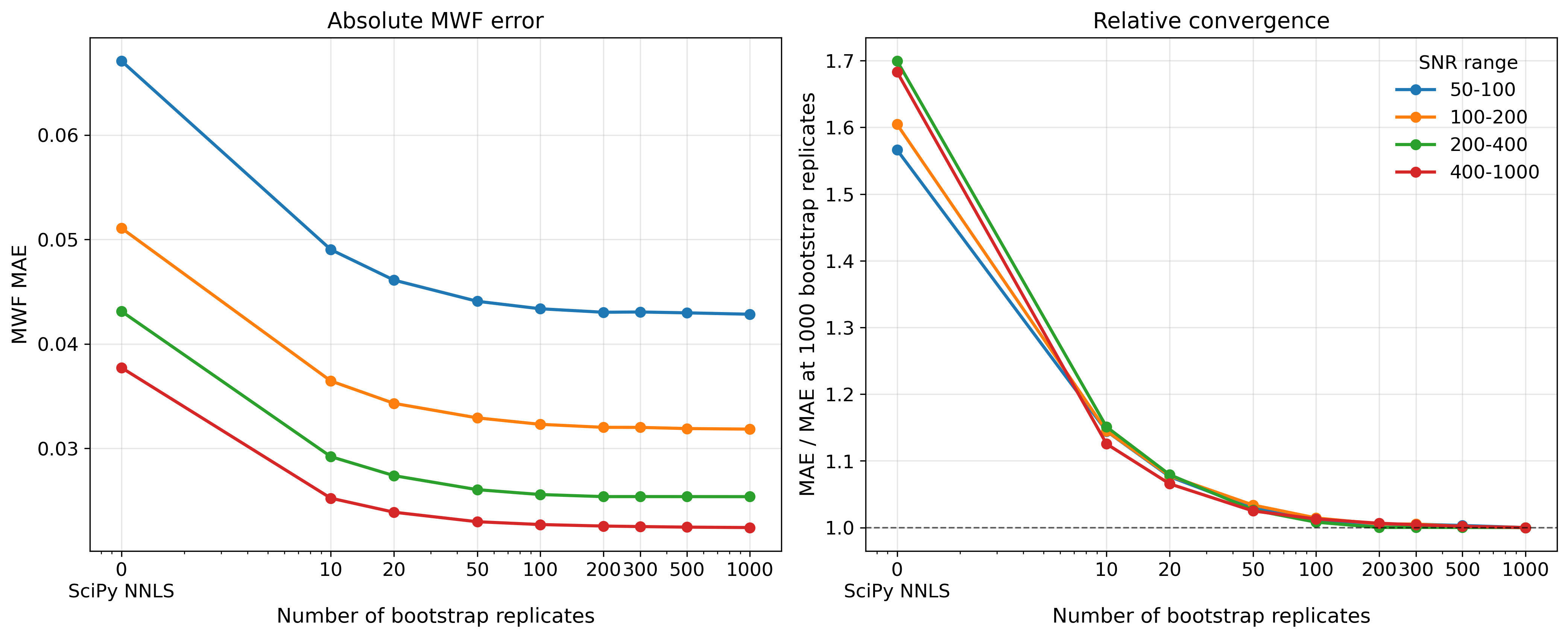}
\caption{
Effect of the number of bootstrap replicates on the mean absolute error (MAE) of the estimated myelin water fraction (MWF). The left panel shows the MWF MAE for each SNR range as a function of the number of bootstrap replicates, from 0 to 1000. 
The case with 0 replicates corresponds to the standard NNLS solution and is included as a reference for the improvement obtained by bootstrap aggregation. The right panel shows the ratio between the MWF MAE obtained with each number of replicates and the value obtained with 1000, providing a normalized measure of convergence toward the largest ensemble size evaluated.
} 
\label{fig1_bootstrap_count}
\end{figure*}

Figure~\ref{fig1_bootstrap_count} shows the effect of the number of bootstrap replicates on the MWF estimation error using \(10{,}000\) simulated voxels per SNR regime. A clear reduction in MAE was already observed with a small number of replicates, indicating that even limited aggregation stabilizes the estimation relative to the standard NNLS solution without aggregation. The improvement was largest for the first replicates and then progressively decreased, with the MAE curves approaching a plateau as the ensemble size increased. Although small additional gains were still observed for larger ensembles, the normalized convergence curves showed that the estimates obtained with 200 bootstrap replicates were already close to those obtained with 1000 for all SNR regimes. Therefore, \(B=200\) was selected as a practical compromise between estimation accuracy and computational burden and was used for the remaining evaluations in this work.

The quantitative performance across methods and SNR regimes is summarized in Table~\ref{tab:multi_snr_metrics}. The primary metrics used to quantify MWF accuracy were MAE and RMSE. According to both metrics, NNLS-BB achieved the best performance in all SNR regimes. The RMSE decomposition into cRMSE and MBE reveals different sources of improvement. Compared with conventional NNLS, NNLS-BB reduced cRMSE by 36–43\% across SNR regimes, indicating less dispersion in MWF estimation errors. Relative to its strongest competitors, Bayes-Reg-\(D^{-1}\) and MIML, its cRMSE was generally comparable, but the average magnitude of its MBE across the four SNR regimes was approximately 82\% and 86\% lower, respectively. Both competing methods consistently underestimated MWF on average, indicating that the advantage of NNLS-BB over these methods was primarily associated with reduced systematic bias.

The remaining metrics provide a complementary view of the reconstruction problem. Bayes-Reg-\(D^{-1}\) and MIML remained highly competitive in terms of correlation \(R\), with MIML giving the highest \(R\) in the intermediate SNR regimes and Bayes-Reg-\(D^{-1}\) among the strongest methods at low SNR. Moreover, MIML consistently achieved the best full-spectrum reconstruction metrics, as measured by RMSE-S and MJSD-S, followed by Bayes-Reg-\(D^{-1}\).

\begin{figure*}[!t]
  \centering
  \includegraphics[width=0.75\textwidth]{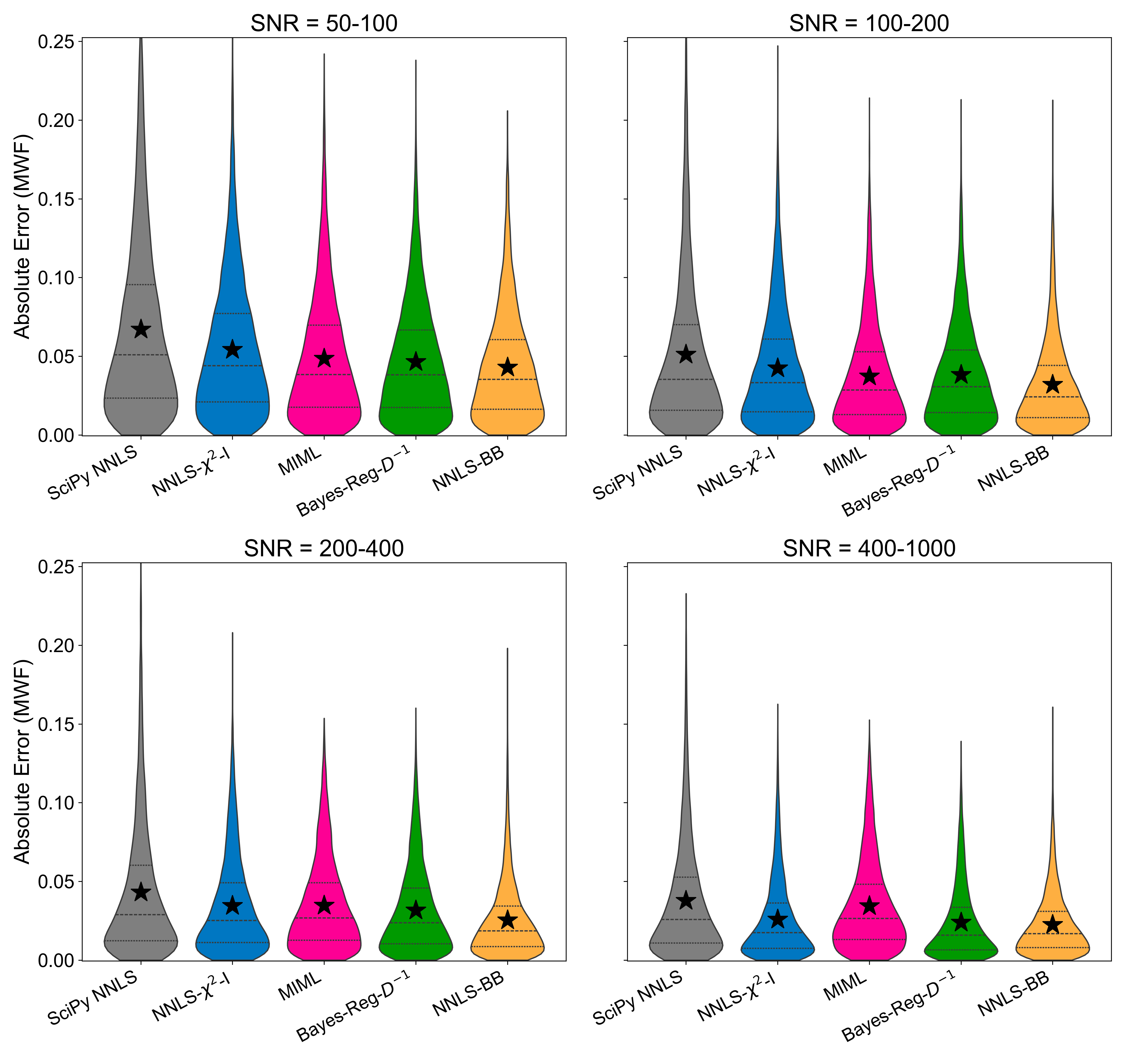}
  \caption{
Voxelwise absolute myelin water fraction (MWF) error distributions across SNR ranges. Each panel corresponds to one SNR interval and summarizes the absolute MWF error over \(10{,}000\) simulated voxels. Violin widths represent the empirical error distribution for each method, internal horizontal markers indicate quartiles, and black stars indicate the mean absolute error. Across all SNR regimes, NNLS-BB shifted the error distribution toward lower values compared with the classical NNLS-based methods and MIML.
}
  \label{fig2_violin_plots}
\end{figure*}

Complementary results are presented in Figure~\ref{fig2_violin_plots}, which shows violin plots of the voxelwise absolute MWF error distributions. Across all SNR regimes, NNLS-BB shifted the error distribution toward lower values compared with the classical NNLS-based methods and MIML. This was reflected in smaller median and mean errors, in agreement with Table~\ref{tab:multi_snr_metrics}, and in more compact interquartile ranges.

The blockwise analysis using 50 independent blocks of 200 voxels confirmed the consistency of the MAE advantage of NNLS-BB. After Holm correction within each SNR regime, NNLS-BB yielded significantly lower blockwise MWF errors in all comparisons
(Wilcoxon signed-rank test, $7.1\times10^{-15}\leq p_{\mathrm{Holm}}
\leq2.6\times10^{-6}$). Lower MAE was observed in $76\%$ to $100\%$
of the blocks, depending on the SNR regime and reference method.

\begin{figure*}[!t]
  \centering
  \includegraphics[width=1\textwidth]{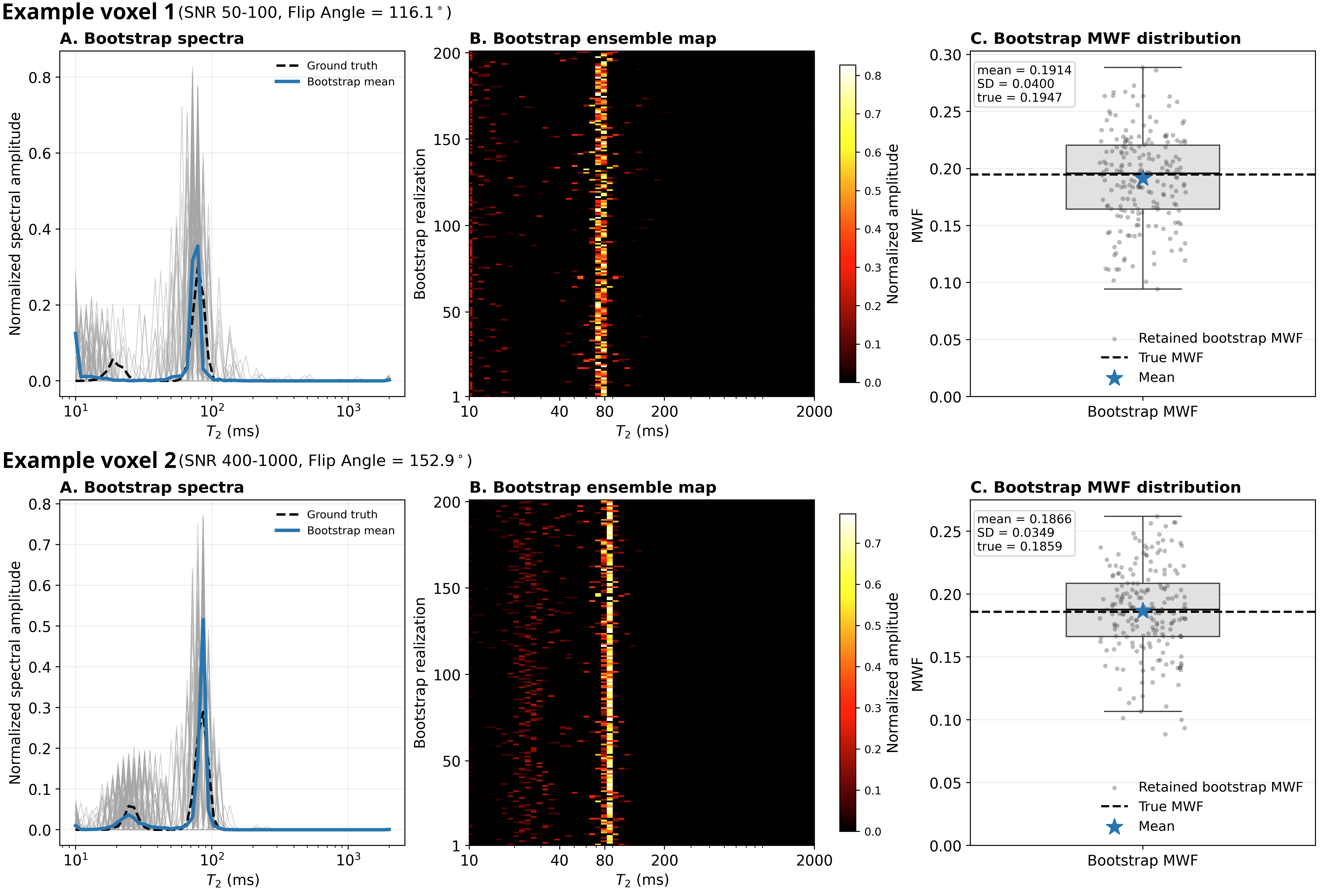}
\caption{
Illustration of the NNLS-BB estimator for two representative voxels. The upper row corresponds to a voxel from the lowest SNR regime (\(\mathrm{SNR}=50{-}100\)), whereas the lower row corresponds to a voxel from the highest SNR regime (\(\mathrm{SNR}=400{-}1000\)). For each voxel, the left panel shows the estimated bootstrap \(T_2\) spectra, the true spectrum, and the bootstrap mean spectrum. The middle panel shows the bootstrap ensemble heat map, where each row corresponds to one bootstrap replicate and each column to one \(T_2\) bin; color encodes the normalized spectral amplitude.
The right panel shows the empirical distribution of the retained bootstrap myelin water fraction (MWF) estimates, together with the bootstrap mean, standard deviation (SD), and ground-truth MWF.
}  
\label{fig3_bootstrap_barycenter}
\end{figure*}

Figure~\ref{fig3_bootstrap_barycenter} illustrates the behavior of NNLS-BB for two representative voxels. The upper row corresponds to a voxel from the lowest SNR regime (\(\mathrm{SNR}=50{-}100\)), whereas the lower row corresponds to a voxel from the highest SNR regime (\(\mathrm{SNR}=400{-}1000\)). For each voxel, the estimated bootstrap spectra are shown together with their bootstrap mean, the corresponding bootstrap ensemble heat map, and the empirical distribution of the retained bootstrap MWF values. In the low-SNR example, the bootstrap mean does not perfectly recover the position of the short-\(T_2\) lobe, which appears shifted toward shorter relaxation times. Nevertheless, the integrated short-\(T_2\) fraction, and therefore the estimated MWF, remains close to the ground truth. This illustrates an important feature of the problem: shifts in the location of the myelin-related lobe do not necessarily imply a large error in the estimated MWF. 
In contrast, in the high-SNR example, the bootstrap mean closely follows the true spectrum. The ensemble heat map further shows that the position and amplitude of the non-dominant short-\(T_2\) component fluctuate more strongly across bootstrap realizations than the dominant intra- and extracellular water component, whose location remains comparatively stable.

\subsection{Real data: scan--rescan repeatability and bootstrap variability}

Table~\ref{tab:retest_metrics} summarizes the scan--rescan
results for the two in vivo cohorts.
In Cohort~1, NNLS-BB achieved the most favorable values
across all reported measures.
In Cohort~2, SciPy NNLS yielded the smallest mean absolute
scan--rescan offset, followed by NNLS-BB, whereas NNLS-BB
achieved the most favorable values for the remaining
measures.
Across both cohorts, NNLS-BB provided the lowest mean
$\mathrm{SD}_{\mathrm{diff}}$, the highest mean correlation,
and regression slopes closest to unity.
The between-subject SD of the correlation coefficients
was also lowest for NNLS-BB.
These results indicate reduced regional scan--rescan
dispersion and stronger correspondence between sessions.

\begin{table*}[!t]
\centering
\caption{
Scan--rescan comparison of MWF estimates in the two
in vivo cohorts.
Metrics were calculated for each subject across the
white-matter tract ROIs and then summarized across subjects.
Mean $|\mathrm{MBE}|$ measures the average magnitude
of the scan--rescan offset, and mean
$\mathrm{SD}_{\mathrm{diff}}$ summarizes the dispersion
of the regional differences.
Mean $r$ and SD $r$ describe the average correlation
between sessions and its variation across subjects.
Mean $|\beta-1|$ measures the deviation of the regression
slope from unity.
Smaller offsets, dispersions, and slope deviations,
together with higher correlations, indicate closer
scan--rescan agreement.
Within each cohort, the most favorable value in each
column is shown in bold and the second most favorable
is underlined.
$S$ denotes the number of subjects.
}
\label{tab:retest_metrics}
\resizebox{\textwidth}{!}{%
\begin{tabular}{llrrrrr}
\toprule
Cohort & Method
& Mean $|\mathrm{MBE}|$
& Mean $\mathrm{SD}_{\mathrm{diff}}$
& Mean $r$
& SD $r$
& Mean $|\beta-1|$ \\
\midrule
Cohort 1 ($S=5$)
& SciPy NNLS
& 0.0057 & 0.0119 & 0.9168 & 0.0311 & 0.1248 \\
& NNLS-$\chi^2$-I
& \underline{0.0039} & 0.0080 & 0.9540 & 0.0261
& \underline{0.0496} \\
& Bayes-Reg-$D^{-1}$
& 0.0042 & \underline{0.0079} & \underline{0.9615}
& \underline{0.0199} & 0.0517 \\
& NNLS-BB
& \textbf{0.0036} & \textbf{0.0073} & \textbf{0.9625}
& \textbf{0.0145} & \textbf{0.0384} \\
\midrule
Cohort 2 ($S=8$)
& SciPy NNLS
& \textbf{0.0023} & 0.0070 & 0.9477 & 0.0260
& \underline{0.0440} \\
& NNLS-$\chi^2$-I
& 0.0029 & 0.0057 & 0.9604 & \underline{0.0253}
& 0.0523 \\
& Bayes-Reg-$D^{-1}$
& 0.0038 & \underline{0.0056} & \underline{0.9676}
& 0.0303 & 0.0458 \\
& NNLS-BB
& \underline{0.0027} & \textbf{0.0054} & \textbf{0.9680}
& \textbf{0.0241} & \textbf{0.0384} \\
\bottomrule
\end{tabular}%
}
\end{table*}

\begin{figure*}[!t]
\centering
\includegraphics[width=\textwidth]
{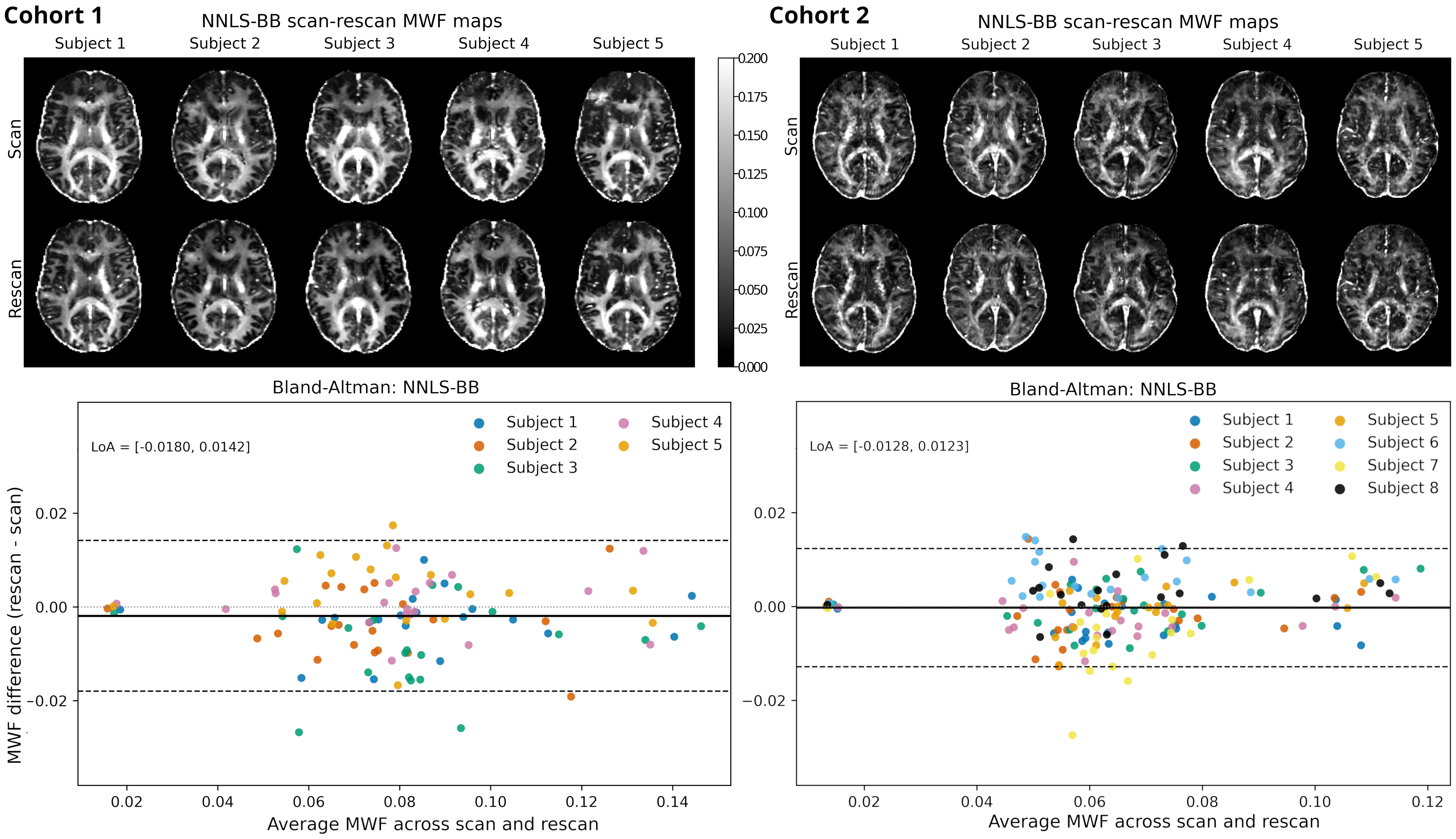}
\caption{
Scan--rescan MWF estimates obtained with NNLS-BB.
The left and right panels correspond to Cohorts~1
and~2, respectively.
The upper panels show axial MWF maps from the scan
and rescan sessions using a common intensity scale.
All five subjects from Cohort~1 are displayed;
five of the eight subjects from Cohort~2 are shown
for a balanced visual layout.
The lower panels show Bland--Altman plots using
regional measurements from all subjects in each
cohort, including all eight subjects in Cohort~2.
Each point represents a subject--ROI pair, with
color identifying the subject.
The horizontal axis shows the average MWF across
sessions, and the vertical axis shows the difference
(rescan minus scan).
Solid black lines indicate the mean differences,
dashed black lines indicate the limits of agreement (LoA), and dotted gray lines mark zero difference.
}
\label{fig:mwf_scan_rescan}
\end{figure*}

Figure~\ref{fig:mwf_scan_rescan} illustrates the
scan--rescan agreement of NNLS-BB at the individual
subject level.
The axial MWF maps show similar spatial patterns
across sessions, with local differences in estimated
amplitudes.
The Bland--Altman plots summarize the corresponding
regional measurements for all subjects.
Cohort~2 showed less scan--rescan dispersion than
Cohort~1, consistent with the lower
$\mathrm{SD}_{\mathrm{diff}}$ values in
Table~\ref{tab:retest_metrics}.
The plotted limits of agreement were
$[-0.0180,\,0.0142]$ for Cohort~1 and
$[-0.0128,\,0.0123]$ for Cohort~2.

\begin{figure*}[!t]
\centering
\includegraphics[width=\textwidth]
{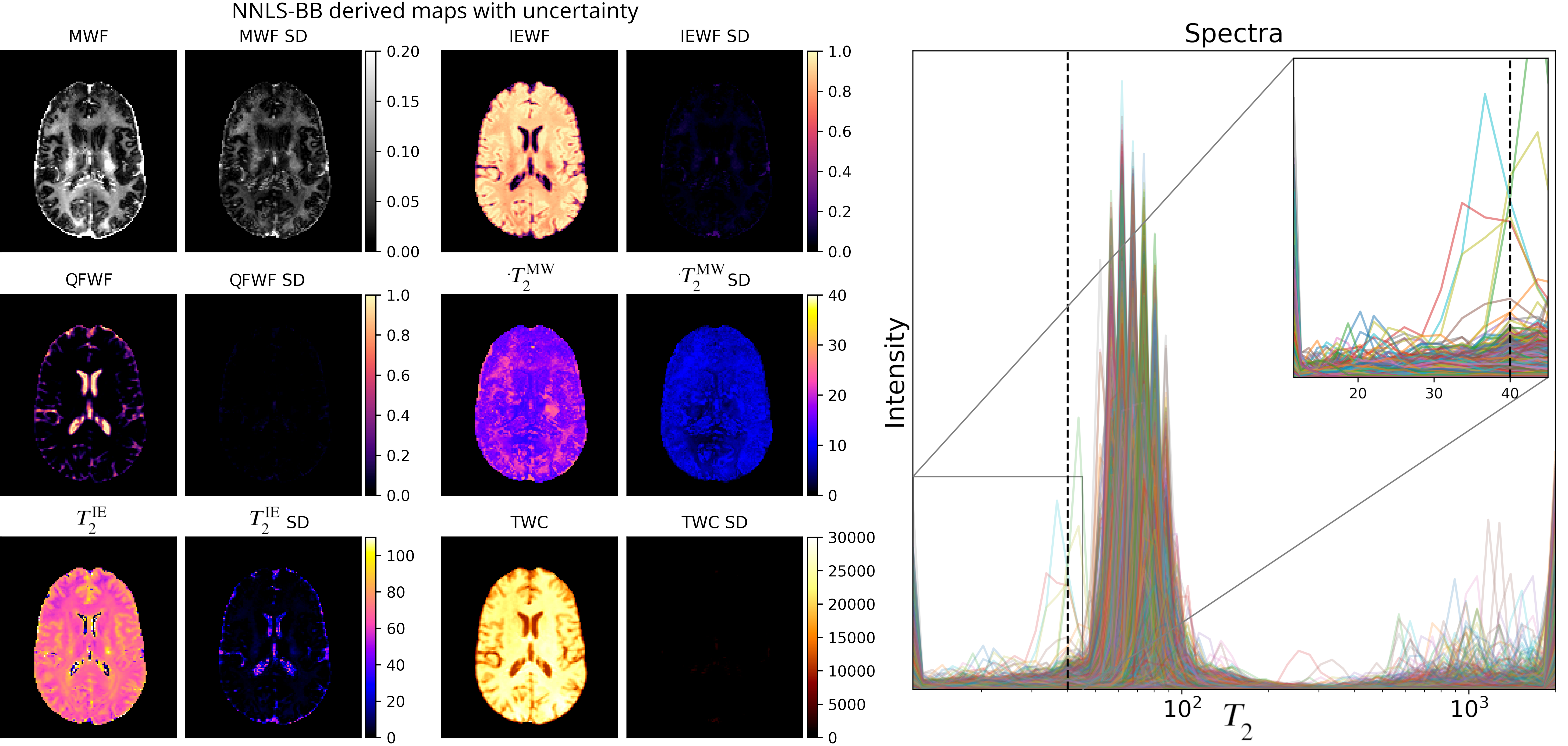}
\caption{
Spectrum-derived parameter maps from NNLS-BB, bootstrap variability, and reconstructed white-matter spectra.
The left panel shows axial maps of MWF, IEWF,
quasi-free water fraction, $T_2^{\mathrm{MW}}$,
$T_2^{\mathrm{IE}}$, and TWC.
Each parameter map is paired with the standard
deviation (SD) of its bootstrap estimates,
using the same color scale within each pair.
The $T_2$ values and their standard deviations
are expressed in milliseconds. TWC and its standard deviation are expressed
in arbitrary signal units.
The right panel shows reconstructed spectra
from voxels within a white-matter mask.
The inset enlarges the short-$T_2$ range,
approximately 10--50~ms.
The dashed vertical line marks the 40~ms cutoff
used to calculate MWF.
}
\label{fig:maps_uncertainty_spectra}
\end{figure*}

Figure~\ref{fig:maps_uncertainty_spectra} shows the
six spectrum-derived parameter maps alongside their
bootstrap standard deviations determined via NNLS-BB.
MWF and $T_2^{\mathrm{MW}}$ exhibited visibly appreciable
bootstrap variability relative to their parameter
scales, consistent with the difficulty of estimating
the smaller short-$T_2$ component. The TWC and QFWF SD maps appeared almost uniformly dark on the color scales shared with their respective parameter maps, indicating very small bootstrap standard deviations relative to the displayed parameter ranges. The IEWF SD map showed a similar pattern, with only small clusters near tissue boundaries. For $T_2^{\mathrm{IE}}$, bootstrap variability was also barely visible within white matter, but was more apparent in ventricular regions and near cerebrospinal fluid--gray matter interfaces.

The spectra extracted from the white-matter mask show dominant intra- and extracellular-water peaks at longer \(T_2\) values, together with smaller myelin-water peaks below the 40 ms cutoff. The enlarged short-\(T_2\) view highlights the voxel-to-voxel variation in these smaller peaks. These spectra illustrate spatial differences in the estimated \(T_2\) distributions, while the corresponding standard-deviation maps characterize the parameter-dependent variability captured by the bootstrap ensemble.

%%%%%%%%%%%%%%%%%%%%%%%%%%%%%%%%%%%%%%%%%%%%%%%%%%%%%%%%%%%%%%%
\section{Discussion}
%%%%%%%%%%%%%%%%%%%%%%%%%%%%%%%%%%%%%%%%%%%%%%%%%%%%%%%%%%%%%%%

This study introduces NNLS-BB, a residual-bootstrap approach that averages repeated NNLS spectra and their derived parameters to obtain bootstrap barycenters, while using the dispersion of the parameter estimates to characterize their variability. NNLS-BB requires neither additional acquisitions nor a training dataset and can be incorporated into existing NNLS reconstruction pipelines. Across a wide range of simulated signals and noise levels, it achieved the lowest MWF MAE and RMSE among the evaluated methods (Table~\ref{tab:multi_snr_metrics} and Figure~\ref{fig2_violin_plots}). The in vivo results also showed favorable regional scan--rescan repeatability in two cohorts acquired with different scanners and protocols (Table~\ref{tab:retest_metrics}). These findings support aggregation as a practical approach to improving MWF estimation without introducing an explicit spectral regularization penalty.

Uncertainty in nuclear magnetic resonance (NMR) relaxation inversion has previously been addressed through several complementary approaches. Bounds on spectrum-derived quantities have been obtained by identifying their extreme values among non-negative spectra satisfying a noise-based misfit criterion \citep{Parker2005AssigningData}. For Tikhonov-regularized reconstruction, analytical variance estimates have been combined with bounds on the bias induced by regularization to characterize uncertainty in spectral amplitudes \citep{Zou2016DeterminingMethod}. Monte Carlo methods have also been used to sample ensembles of spectra under a Gaussian noise model and specified prior assumptions, yielding mean spectra and distributions of derived quantities \citep{Prange2009QuantifyingInversion, Prange2010UnderstandingUncertainty}.

Resampling has also been applied to surface NMR, where repeated fitting to randomly selected subsets of the measured data was used to characterize variability in water-content and relaxation-time profiles while retaining the full-data inversion as the point estimate \citep{Parsekian2015UncertaintyStatistics}. NNLS-BB instead resamples fitted residuals and repeatedly solves the NNLS problem, using the resulting ensemble both to obtain the final parameter estimates through aggregation and to describe their bootstrap variability.

Aggregation has also been used in MWI to stabilize reconstructions obtained with orthogonal matching pursuit, a greedy algorithm whose results can depend strongly on initialization, even for identical input data \citep{Drenthen2019APursuit}. This approach combines repeated reconstructions from random initializations using residual-dependent weights to reduce the influence of poorer fits on the final estimate. In contrast, NNLS-BB generates an ensemble through residual-based signal perturbations, addressing variability associated with measurement noise rather than algorithm initialization, while also providing voxelwise estimates of bootstrap variability. More broadly, noise perturbation and ensemble averaging have been studied as mechanisms for stabilizing estimators \citep{Raviv1996BootstrappingTechnique}, providing further motivation for this approach.

The geometric analysis developed in Section~\ref{sec:theory} helps interpret our results. Rather than selecting a single spectrum among nearly equivalent solutions, NNLS-BB combines reconstructions obtained under different noise perturbations, reducing the influence of spectral fluctuations that vary across the ensemble (Figure~\ref{fig3_bootstrap_barycenter}). Because the non-negativity constraint makes NNLS nonlinear in the measured data, the average of these reconstructions does not generally coincide with the original NNLS solution. The simulations suggest that this strategy can reduce error dispersion relative to conventional NNLS while exhibiting substantially less systematic bias than regularized methods under the evaluated conditions (Table~\ref{tab:multi_snr_metrics}).

An important finding is that improved MWF estimation did not require the most accurate reconstruction of the full spectrum. MIML achieved the best spectral reconstruction metrics, whereas NNLS-BB achieved the lowest MWF MAE and RMSE (Table~\ref{tab:multi_snr_metrics}). Spectral metrics assess discrepancies across the entire distribution, while MWF depends on the relative signal contribution of water components with short $T_2$ values. This distinction is also illustrated in Figure~\ref{fig3_bootstrap_barycenter}, where the low-SNR example yields an MWF estimate close to the ground truth despite a displacement of the short-$T_2$ lobe. These results support evaluating methods directly in terms of the parameter of interest.

The MIML results should be interpreted in relation to the training conditions. We expanded the training SNR range from the original $80$--$200$ \citep{Yu2020Model-InformedRelaxometry} to $50$--$500$, while the evaluation extended to an SNR of $1000$. Part of the highest evaluation interval therefore lay outside the training range, and training specifically adapted to higher SNRs could improve its performance in that interval. Nevertheless, NNLS-BB also achieved lower MWF MAE and RMSE in the three lower SNR intervals, which were fully contained within the MIML training range (Table~\ref{tab:multi_snr_metrics}).

While the simulations assessed estimation accuracy against known ground truth, the in vivo experiments provided complementary evidence of scan–rescan repeatability. Notably, NNLS-BB yielded the lowest mean regional scan–rescan dispersion in both cohorts, together with high correlations and regression slopes close to unity (Table~\ref{tab:retest_metrics}). The individual maps and Bland--Altman plots further illustrate the scan--rescan agreement (Figure~\ref{fig:mwf_scan_rescan}). The consistency across two acquisition protocols supports its applicability beyond a single dataset. However, the cohorts were small and consisted of young healthy participants. Larger studies including pathological tissue are needed to establish clinical applicability. Because different participants were examined in the two cohorts, the experiments assess within-protocol repeatability. Cross-scanner agreement remains to be evaluated, given the potential influence of acquisition differences \citep{Fischi-Gomez2022VariabilityAcquisitions}.

The bootstrap-variability maps complement the parameter estimates by showing their sensitivity to signal perturbations (Figure~\ref{fig:maps_uncertainty_spectra}). MWF and myelin-water $T_2$ showed visible variability, whereas the other parameters generally had much lower SD values relative to their displayed ranges. In future clinical research, this information could help assess the stability of regional MWF abnormalities or longitudinal changes, including measurements within lesions. It could also support quality assessment and comparisons of acquisition protocols in terms of the precision of the parameters of interest. A further potential application is to incorporate estimation variability into population-level statistical analyses. Alongside the mean parameter maps, SD maps could inform group comparisons and regression analyses that account for differences in estimation precision across subjects. Related approaches have been demonstrated in diffusion MRI \citep{Sjolund2018BayesianMRI}.

The main practical cost of NNLS-BB is the repeated spectral fitting. In the current implementation, it was slower than the other evaluated reconstruction methods (Table~\ref{tab:multi_snr_metrics}). The ensemble-size analysis revealed diminishing improvements as the number of replicates increased, with $B=200$ producing estimates close to those obtained with substantially larger ensembles (Figure~\ref{fig1_bootstrap_count}). This provided a practical compromise between accuracy and computation time. Smaller ensembles, for example around $B=100$, or adaptive selection of the ensemble size could offer further computational savings where a modest loss of estimation accuracy is acceptable.

Overall, the results support NNLS residual-bootstrap aggregation as a useful approach to MWF estimation, combining lower errors in simulations with favorable scan--rescan repeatability and complementary parameter-variability maps in vivo. Further work should prioritize evaluation in clinical populations.

\section*{CRediT authorship contribution statement}

\textbf{E.J.C.-R.:} Supervision, Conceptualization, Investigation, Methodology, Project administration, Data curation, Formal analysis, Resources, Software, Validation, Visualization, Writing -- original draft, Writing -- review \& editing.
\textbf{T.Y.:} Investigation,  Data acquisition, Methodology, Software,
Writing -- review \& editing.
\textbf{J.M.G.:} Investigation, Resources, Writing -- review \& editing.
\textbf{S.G.-Á.:} Investigation, Data acquisition, Resources, Writing -- review \& editing.
\textbf{T.K.:} Investigation, Data acquisition, Resources, Writing -- review \& editing.
\textbf{T.H.:} Investigation, Data acquisition, Data curation, Writing -- review \& editing.
\textbf{G.F.P.:} Investigation, Data acquisition, Data curation, Writing -- review \& editing.
\textbf{M.A.R.-F.:} Investigation, Data acquisition, Data curation, Resources, Supervision, Writing -- review \& editing.

\section*{Data and code availability}

The reconstruction code is publicly available at
\url{https://github.com/ejcanalesr/multicomponent-T2-toolbox}.
The human MRI data are not publicly available due to
ethical and privacy restrictions. Requests for access
may be considered subject to the applicable consent
conditions and institutional approvals.

\section*{Funding}
\begin{sloppypar}
E.J.C.-R. acknowledges support from the Ramón y Cajal fellowship RYC2023-042763-I, funded by the Spanish Ministry of Science, Innovation and Universities and the Spanish State Research Agency (MICIU/AEI/\allowbreak10.13039/\allowbreak501100011033), and co-funded by the European Social Fund Plus (ESF+). Data acquisition for Cohort~2 was supported by grants from the Ministerio de Economía y Competitividad, Spain
(PSI2014-55316-C3-3-R, PSI2017-89389-C2-2-R, and PID2020-\allowbreak114521RB-\allowbreak C21/C22), and the Consellería de Cultura, Educación e Ordenación Universitaria, Xunta de Galicia
(ED431-2017/27 and ED431C-2021/04), all co-funded by the European Regional Development Fund (ERDF).
\end{sloppypar}

\section*{Declaration of competing interest}
G.F.P., T.H., and T.Y. are employees of Siemens Healthineers International AG, Switzerland. T.K. is an employee of Siemens Healthineers AG, Germany. The remaining authors declare no competing interests.

\section*{Declaration of generative AI and AI-assisted
technologies in the manuscript preparation process}

During the preparation of this work, the authors used ChatGPT to assist with drafting and organizing manuscript text and improving its clarity, grammar, and readability. The authors reviewed and edited the resulting text, verified the scientific content and references, and take full responsibility for the content of the published article.

%% Loading bibliography style file
\bibliographystyle{plainnat}
\bibliography{references}

\end{document}